# A Wideband Narrow Beam 1x6 Linear Antenna Array for Automotive Radar and 5G Millimetre-Wave Applications

Muhammad Asfar Saeed
*School of Engineering*
*Faculty of Engineering & Science*
*University of Greenwich*
Kent, UK
m.a.saeed@greenwich.ac.uk

Augustine O. Nwajana
*School of Engineering*
*Faculty of Engineering & Science*
*University of Greenwich*
Kent, UK
a.o.nwajana@greenwich.ac.uk

***Abstract*—This paper presents the design and performance analysis of a 1×6 linear microstrip patch antenna array tailored for automotive radar and 5G millimetre-wave (mm-wave) applications. The proposed antenna array comprises six rectangular radiating patches with the primary patch excited using a microstrip feedline, while the remaining patches are interconnected through narrow microstrip lines with a width of 0.1 mm, enabling effective power distribution along the array. Optimal inter-element spacing facilitates constructive and destructive interference, enabling the formation of a narrow beam with enhanced directivity and a wide operational bandwidth. The high-gain radiation characteristics are achieved through the combined effects of the six-element linear configuration and precise impedance matching. Key performance metrics including reflection coefficient, current distribution, and radiation patterns have been analysed. Results demonstrate a reflection coefficient better than 10 dB across the target frequency range and a narrow beamwidth with high directivity, making the array suitable for high-resolution automotive radar and 5G mm-wave communications. Potential applications include vehicle-to-vehicle (V2V) radar sensing, lane change detection, blind spot monitoring at 28 GHz, and high-capacity point-to-point wireless backhaul links. The design offers a promising solution for compact, high-performance beamforming antenna systems in intelligent transportation and next-generation wireless networks.**

## I. Introduction

Millimetre-wave (mm-wave) frequency bands, typically ranging from 24 to 300 GHz, have gained significant attention in recent years due to their potential for supporting high-capacity, low-latency communication systems. These bands are critical for a wide array of applications, including 5G cellular networks, wireless Gigabit, vehicular communication systems, radar imaging, satellite links, and emerging medical technologies [1]–[5]. The primary advantage of mm-wave communication lies in the availability of broader spectral bandwidths, which enables ultra-fast data rates and supports dense user environments, as illustrated in Fig. 1.

However, mm-wave signals suffer from more pronounced small-scale fading compared to traditional microwave bands. This degradation primarily results from increased multipath propagation as the number of reflecting and absorbing objects in the environment rises, leading to severe signal attenuation due to high free-space path loss and multi-object absorption [1], [6], [7]. Although several mm-wave antenna arrays have been reported in the literature, many rely on multilayer substrates, SIW-based feeds, aperture-coupling, parasitic loading, or MIMO architectures to achieve high gain or wide bandwidth. In contrast, the novelty of this work lies in the development of a compact, single-layer 1×6 series-fed microstrip array that achieves narrow-beam radiation, high directivity, and wide impedance bandwidth at 28 GHz using a simple planar structure. The use of ultra-narrow 0.1 mm microstrip interconnects and carefully optimized 1.9 mm spacing minimizes mutual coupling while preserving wideband behavior as a known challenge in mm-wave series-fed arrays. Furthermore, unlike many prior works, the proposed antenna is experimentally validated and physically fabricated using a low-cost milling process, demonstrating practical suitability for automotive radar and 5G mm-wave applications. To mitigate these challenges, system designers typically rely on either increasing the transmit power or enhancing antenna gain to improve the effective isotropic radiated power (EIRP). In many scenarios, however, especially in automotive and medical environments, there are strict regulations on electromagnetic field (EMF) exposure, limiting the allowable transmit power. In such cases, increasing antenna gain becomes a practical and efficient alternative to boost EIRP. Since antenna gain is directly related to directivity, which is inversely proportional to beamwidth, the use of high-gain narrow-beam antennas becomes essential. As the size and element count of an antenna array increase, the resulting beam narrows, which improves spatial resolution, increases channel capacity, and reduces interference, while maintaining a wider operational bandwidth.

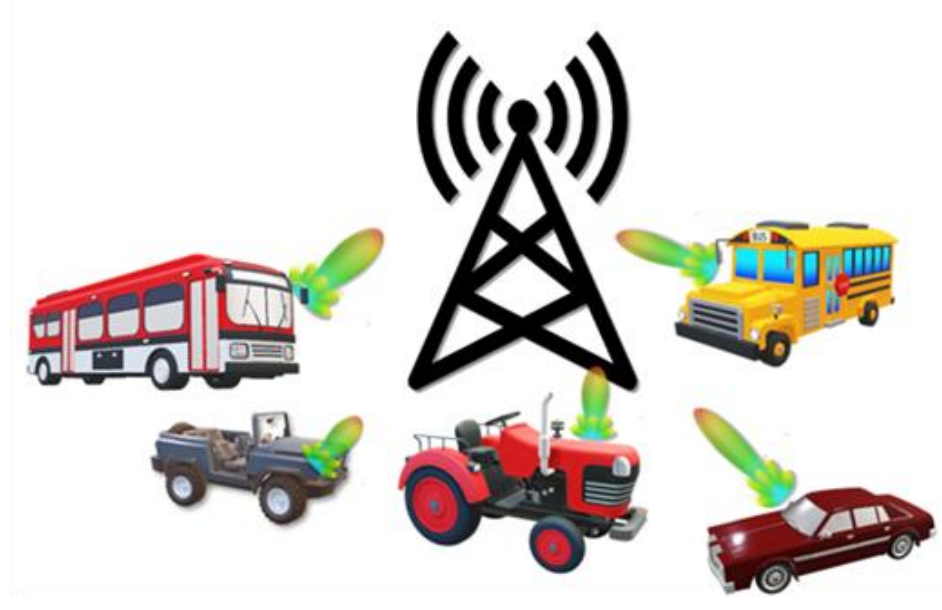

Fig. 1. Wideband antenna array applications in vehicles and 5G network communications.

Directional radiation beams can be synthesized using antenna arrays arranged in one-dimensional (1D) or two-dimensional (2D) configurations. A 1D linear array typically forms a fan-shaped beam in the azimuth or elevation plane, offering a narrow beamwidth in the direction of the array axis and a broader beamwidth in the orthogonal plane. This pattern is well-suited for applications such as automotive radar, mobile devices, and medical imaging systems [8]–[10]. In contrast, 2D planar arrays generate pencil beams, making them ideal for base station deployment and satellite communication links. Various array antenna designs and beamforming techniques have been explored in the literature, including leaky-wave antennas [11], traveling-wave structures [12], magneto-electric (ME) dipole arrays [13], Luneburg lens-based designs [14], and planar arrays with diverse feeding networks [15]–[17]. Despite these advancements, there remains a strong demand for compact, low-profile, high-gain antenna arrays with simple feed mechanisms and broad bandwidth for mm Wave applications. For instance, the leaky-wave antenna (LWA) described in [11] employs a series-fed configuration to minimize losses within the feed network; however, this approach inherently limits the operational bandwidth. To overcome this limitation, a microstrip-to-slot transition technique was introduced, which significantly enhances bandwidth performance. In [18], an LWA based on a half-mode substrate-integrated waveguide (HMSIW) aperture was proposed, offering improved bandwidth and end-fire radiation with a pencil beam by optimizing the aperture width. Nevertheless, the use of metallic vias to emulate electric walls in HMSIW increases fabrication complexity. This was addressed in [19] through a planar open-circuited corrugated design that eliminates the need for vias. In array antenna design, sidelobe level (SLL) suppression is a critical challenge. The study in [15] achieved effective SLL reduction by employing a series-fed array composed of non-uniformly shaped radiating elements. Similarly, integrating electromagnetically coupled split-ring resonators led to reduced SLL in [20], although this configuration introduced beam squint. This drawback was mitigated using log-periodic antenna structures, as demonstrated in [21], [22] which maintained low SLL while avoiding beam squint.

An alternative to series feeding is the parallel feeding network, which enables independent control of the power and phase delivered to each radiating element [8]. For example, in [23], an 8-element patch antenna utilizing substrate-integrated waveguide (SIW) feeding and a pair of I-shaped parasitic elements was proposed to generate an elliptic radiation beam. Meanwhile, [24] introduced a directional beam in both the XZ- and YZ-planes using two pairs of transverse slots etched on a metallic layer, with the SIW feed located on the bottom layer. Despite its advantages, SIW feeding is inherently complex, requiring meticulous spacing and precision milling to form effective electromagnetic walls and prevent leakage.

To simplify the feeding architecture, [25] proposed a planar parallel feed structure using a Bezier-curved bowtie antenna array. Additionally, in [26], both an 8-element linear array and a 2×4 array were implemented with parallel feeding networks. The study also introduced a hybrid feeding mechanism for the 2×4 configuration. This hybrid approach produced unequal power distribution across the array elements, which effectively enhanced the operational bandwidth and achieved high gain performance at 60 GHz A hybrid feeding approach, as demonstrated in [27], integrates both series and parallel feeding methods to extend the operational bandwidth. However, this configuration introduces non-uniform power distribution among the elements, resulting in a distorted radiation pattern. In [28], a centre-fed architecture employing Taylor power distribution was implemented to improve pattern control. Nonetheless, the design exhibited limited bandwidth. Alternatively, [29] employed Chebyshev-based unequal power distribution to effectively suppress SLL. While this method achieved low SLL, the bandwidth remained narrow, confined to the 23.5–24.5 GHz range.

The reviewed literature highlights various design approaches aimed at achieving high gain, wide bandwidth, and narrow beam characteristics in antenna arrays. However, many of the reported designs, particularly leaky-wave antennas (LWAs) and multilayer structures employing SIW-based feed networks, tend to increase structural complexity and overall antenna profile [29]. In contrast, this article proposes a compact 1×6 linear microstrip patch antenna array that incorporates the desired features while maintaining a simple, low-profile configuration suitable for K- and Ka-band applications. Nevertheless, mutual coupling among closely spaced elements across the wide bandwidth may introduce non-uniform power distribution, potentially leading to beam squint and radiation asymmetry. The proposed antenna is well-suited for a range of modern high-frequency applications, including automotive radar, vehicle-to-vehicle (V2V) and vehicle-to-everything (V2X) communication, point-to-point wireless links, next-generation healthcare monitoring systems, 5G/6G cellular networks, satellite communications, and compact Internet of Things (IoT) devices. Its wideband operation and high gain characteristics make it a strong candidate for integration into multi-functional communication platforms [30], [31]. The antenna can be interfaced with multiple on-board transceivers used in automotive radar and communication systems. By incorporating a bank of bandpass filters to mitigate adjacent-band interference and a high-isolation multiplexer for dynamic transceiver selection, the design supports seamless operation across various mm-wave frequency bands [25], [32]. Although series-fed microstrip patch antenna arrays have been widely investigated, achieving the combined characteristics of compact size, wide impedance bandwidth, high gain, and narrow-beam radiation at mm-wave frequencies using a simple planar architecture remains a challenge. Most reported high-gain mm-wave arrays rely on complex multilayer structures, substrate-integrated waveguide (SIW) feeds, corporate feed networks, or leaky-wave configurations, which increase fabrication complexity, cost, and profile thickness.

In contrast, the novelty of this work lies in the realization of a compact, single-layer 1×6 series-fed microstrip patch antenna array operating at 28 GHz that simultaneously achieves wide impedance bandwidth, narrow-beam radiation, and high directional gain using an ultra-simple feeding topology. The optimized inter-element spacing and narrow transmission-line interconnections enable controlled current distribution, reduced mutual coupling, and stable radiation performance across the operating band. To the best of the authors' knowledge, this combination of wideband performance, narrow-beam radiation, and simple planar fabrication for a 1×6 series-fed array specifically targeted at both automotive radar and 5G mm-wave applications has not been previously reported. This enables reliable and efficient frequency reuse, spectrum agility, and real-time beam switching, which are essential for dense vehicular environments, low-latency healthcare telemetry, and massive IoT deployments. The antenna's compact and planar form factor further enhances its suitability for integration into space-constrained platforms such as autonomous vehicles, wearable devices, and drone-based communication relay.

## II. Methodology

The proposed antenna consists of a linear microstrip array composed of six rectangular radiating patches. These patches are arranged in a straight line along a rectangular substrate, forming a 1×6 linear array as shown in Fig. 2. Each patch element is rectangular in shape and interconnected via a narrow transmission line, forming a series-fed structure. This configuration facilitates constructive interference in the broadside direction, contributing to high directivity and narrow beamwidth suitable for mm-wave applications. The antenna is designed on Rogers RO3003 substrate, selected for its low dielectric loss and stable performance at mm-wave frequencies. The substrate has a thickness of 1.574 mm, a dielectric constant ($\varepsilon r$) of 3.0, and a loss tangent ($\tan \delta$) of 0.0013. These properties contribute to minimal signal attenuation and enhanced radiation efficiency.

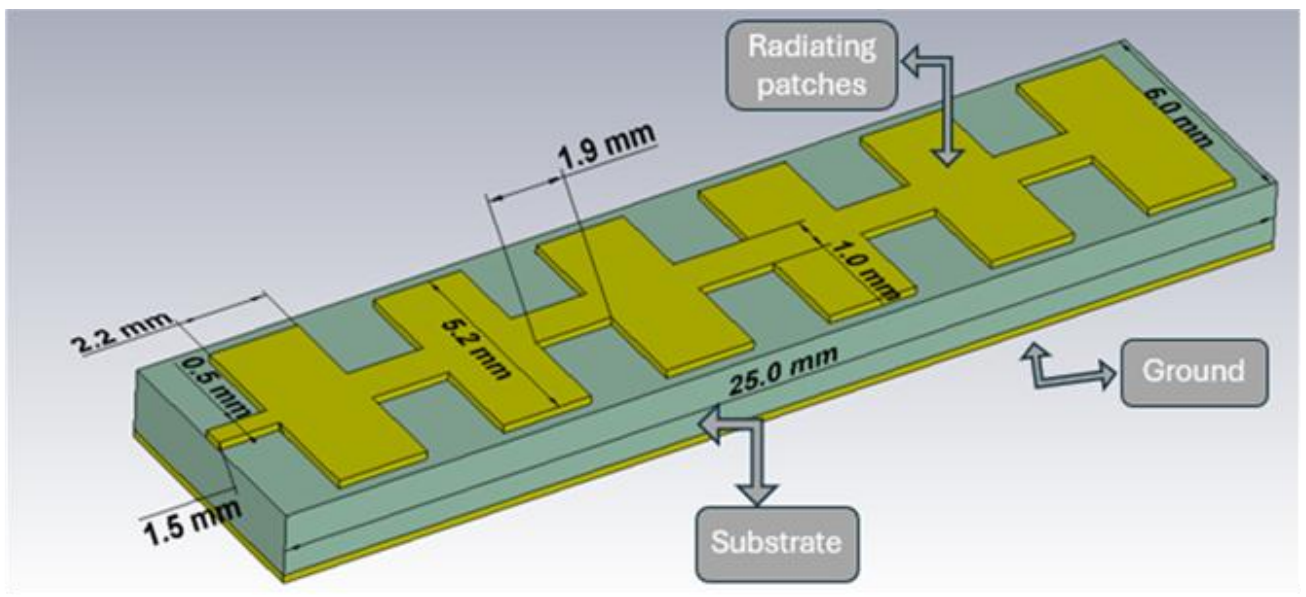

Fig. 2. Layout of the proposed 1x6 linear antenna array.

The antenna array employs a microstrip line feeding method, which offers a compact, low-profile solution that is easy to fabricate and suitable for integration with printed circuit boards. This method also supports precise impedance matching and can be readily optimized for various high-frequency applications. The first radiating patch is directly excited via a microstrip line feed, with dimensions of 0.5 mm in width and 1.5 mm in length. The subsequent patches in the array are interconnected using narrow microstrip transmission lines, each having a width of 0.5 mm and a length of 1.9 mm. This series-fed arrangement ensures a progressive distribution of excitation across all elements, promoting phase coherence and enabling the formation of a high-gain, narrow beam in the broadside direction. The spacing between adjacent radiating elements in the linear array is maintained at 1.9 mm. This inter-element spacing is carefully optimized to minimize mutual coupling effects and to ensure constructive interference of the radiated fields. Such spacing contributes to enhanced antenna gain and the formation of a directive beam with a narrow beamwidth. Additionally, the spacing supports the broadening of operational bandwidth by maintaining phase coherence across the array, which is essential for achieving high performance in millimetre-wave applications. The antenna design was simulated using CST Studio Suite, selected for its user-friendly interface, advanced visualization capabilities, and efficient mesh computation. CST provides accurate full-wave electromagnetic simulation results, making it ideal for evaluating high-frequency antenna structures. The frequency range considered for simulation is from 25 GHz to 35 GHz, targeting operation within the K- and Ka-bands, which are relevant for 5G and automotive radar applications. A time-domain solver was employed due to its suitability for broadband antenna analysis and its capability to efficiently compute S-parameters and radiation characteristics over a wide frequency range. Open (add space) boundary conditions were applied to replicate free-space radiation behaviour, ensuring accurate far-field calculations and minimal reflection artifacts during simulation.

The proposed antenna was fabricated using an electronic milling machine, where the simulated design dimensions were fed directly into the milling software to etch the copper layer on the Rogers 3003 substrate. After the automated milling process, minor copper irregularities or excess material were manually corrected using a half-round copper file to ensure clean patch outlines and proper inter-element spacing. To enable signal feeding and testing, an SMA connector was carefully soldered to the transmission line using a thin soldering iron, ensuring a precise and low-loss connection. This fabrication approach allowed the physical realization of the antenna with high accuracy, closely matching the simulated model. The fabricated model, under test set-up in an anechoic chamber, is shown in Fig. 3.

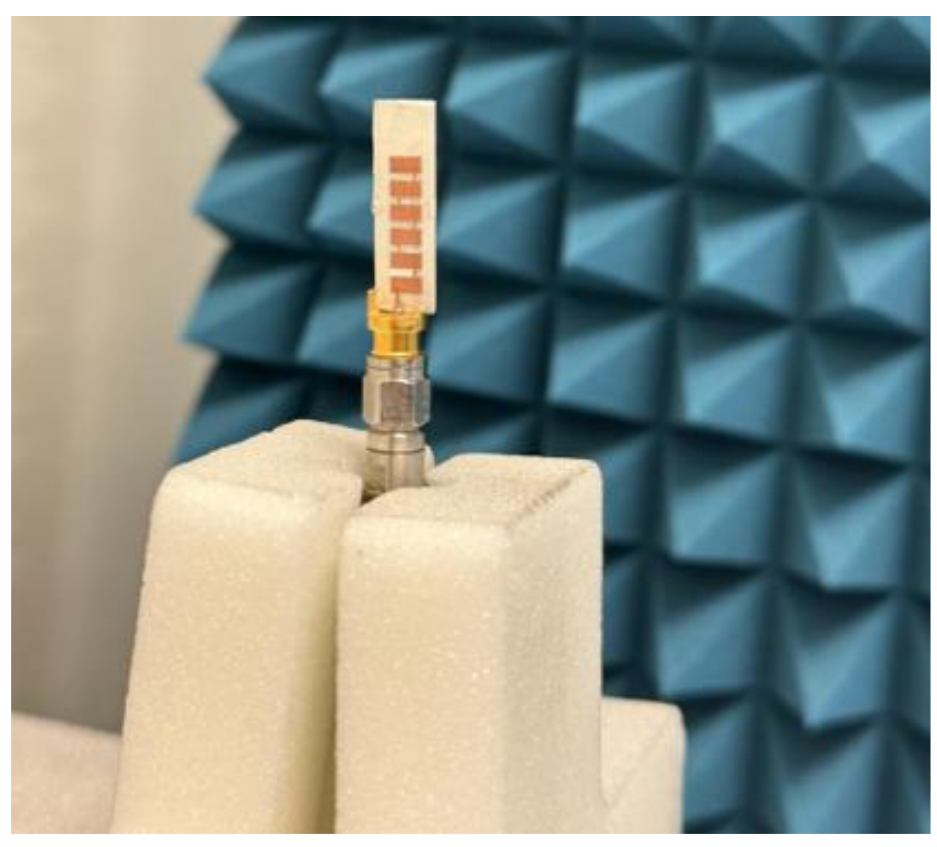

Fig. 3. Fabricated prototype of the 1x6 linear antenna array under test.

## III. Results and Discussion

The proposed antenna array was evaluated based on several key performance parameters critical to high-frequency applications. At the target frequency of 28.5 GHz, the simulated reflection coefficient was measured at approximately 18.0 dB, indicating efficient impedance matching to a standard 50 Ω system. The Voltage Standing Wave Ratio (VSWR) remained well below 2.0, confirming minimal reflection and efficient power transfer. The antenna exhibits a wide impedance bandwidth of 1.2 GHz, covering a significant portion of the K- and Ka-bands.The measured −10 dB impedance bandwidth extends from 27.5 GHz to 28.7 GHz, corresponding to a bandwidth of approximately 1.2 GHz, which is suitable for 28 GHz 5G mm-wave and automotive radar applications. A peak gain of 9.91 dBi was observed, supporting its suitability for high-directivity applications such as automotive radar and 5G mm-wave communication.The proposed antenna achieves a peak realized gain of approximately 9.9 dBi at 28 GHz, along with good radiation efficiency and narrow-beam characteristics. The measured results show good agreement with simulations, confirming the reliability of the proposed design for practical 5G mm-wave and automotive radar applications.The far-field radiation pattern demonstrated a well-defined narrow beam directed along the broadside, with low sidelobe levels. Current distribution analysis revealed that the current flows uniformly from the primary feed point along the series-connected radiating patches, forming standing waves that contribute to consistent radiation across the array. It is observed that the main-beam directions in the $\phi = 0°$ and $\phi = 90°$ planes are not identical, indicating a moderate beam tilt. Such behavior is commonly observed in wideband series-fed antenna arrays and arises from frequency-dependent phase progression and slight non-uniform excitation of the array elements. Despite this beam tilt, the antenna maintains a stable narrow-beam radiation characteristic with well-suppressed sidelobes, which is acceptable for the intended 28 GHz automotive radar and 5G mm-wave applications. In the $\varphi = 0°$ plane, the main beam is steered toward 40.0°, with a slightly wider 3 dB beamwidth of 86.3° and a higher directivity of 4.82 dBi. The sidelobe level in this case is –3.1 dB. The radiation pattern of the proposed antenna array exhibits a directional behavior that is essential for point-to-point 5G mm-wave and automotive radar applications. This behaviour, while common in series-fed wideband antennas, is an important consideration when targeting consistent beam alignment over the operational frequency range.

The reflection coefficient plot illustrates good impedance matching at the target frequency of 28.8 GHz, where a deep null of approximately –33 dB is observed. This indicates that minimal power is reflected to the source, ensuring efficient radiation from the antenna.The proposed antenna exhibits a measured −10 dB impedance bandwidth of 1.2 GHz (27.5–28.7 GHz) with good impedance matching at 28 GHz, confirming its suitability for mm-wave communication systems This wideband behaviour is well suited for modern mm-wave communication systems. Complementing this, the VSWR plot demonstrates values well below 2.0 across the same frequency range, confirming good impedance matching. The minimum VSWR achieved is close to 1.06 at 28.8 GHz, which signifies very good ideal matching at resonance.A slight frequency shift is observed between the simulated and measured reflection coefficient responses. This discrepancy can be attributed to several practical factors, including fabrication tolerances introduced during the PCB milling process, minor variations in the dielectric constant of the Rogers RO3003 substrate, and parasitic effects associated with the soldered SMA connector. In addition, measurement uncertainties and cable losses at mm-wave frequencies can also contribute to this shift. Such variations between simulation and measurement are common in high-frequency antenna implementations and do not significantly affect the overall performance trend of the proposed design. These results validate that the proposed antenna is highly efficient within the designated frequency band. The simulation and measured reflection coefficient results are presented in Fig. 4.

The radiation pattern of the proposed antenna array at the operating frequency of 28 GHz for two principal azimuth planes, namely $\phi = 0°$ and $\phi = 90°$, which are used to evaluate the directional and beamwidth characteristics of the antenna is shown in Fig. 5 and exhibits a directional behaviour which is essential for point-to-point communication in 5G mm-wave applications. At $\phi = 0°$, the main lobe points towards $\theta \approx 15°$, with a peak magnitude of approximately 10 dBi and a 3 dB angular beamwidth of about 86.3°. At $\phi = 90°$, the main beam is also narrow and highly directive, with a comparable 3 dB beamwidth, while the sidelobes remain well suppressed, as illustrated in Fig. 5. These results confirm that the proposed 1×6 linear array provides a stable, narrow-beam radiation pattern with high directivity at 28 GHz. It should be noted that the term "narrow beam" in this work refers to the directional fan-beam radiation characteristic of a one-dimensional linear array. Compared with a single microstrip patch or smaller linear arrays, the proposed 1×6 configuration significantly narrows the radiation beam in the plane of the array, resulting in improved directivity and angular resolution.

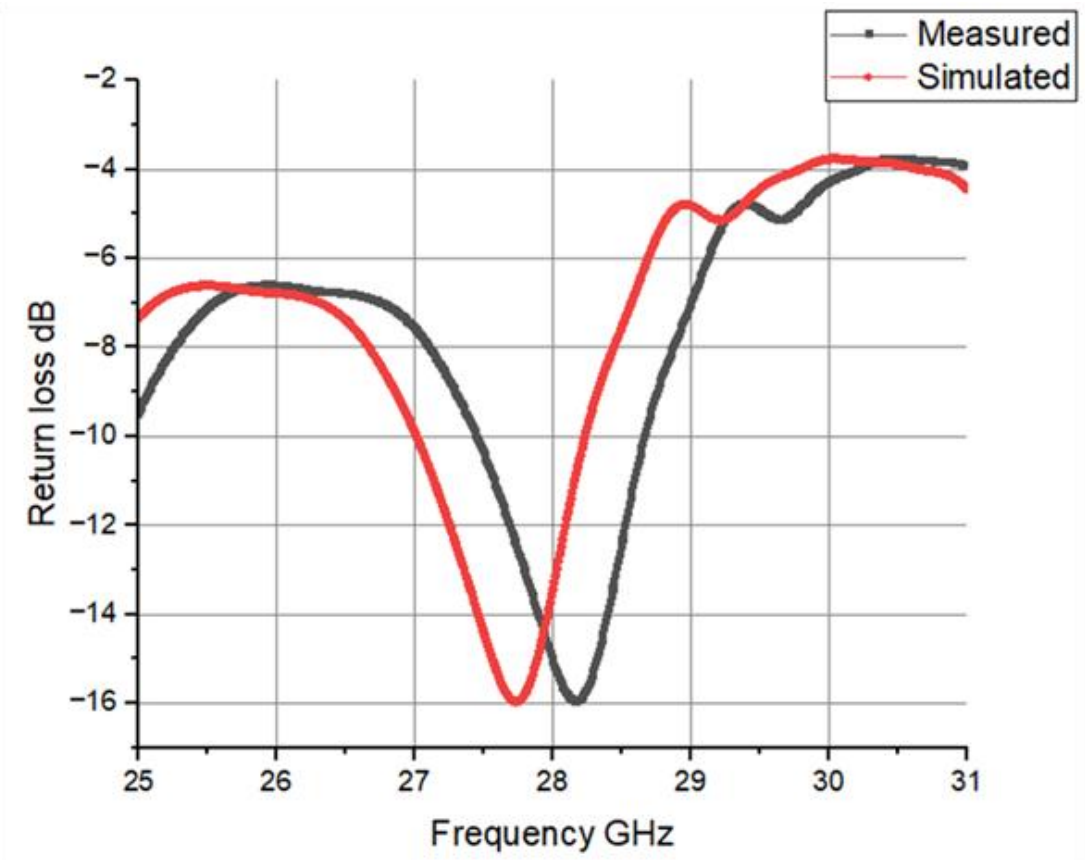

Fig. 4. Reflection coefficients of the proposed 1x6 linear antenna array.

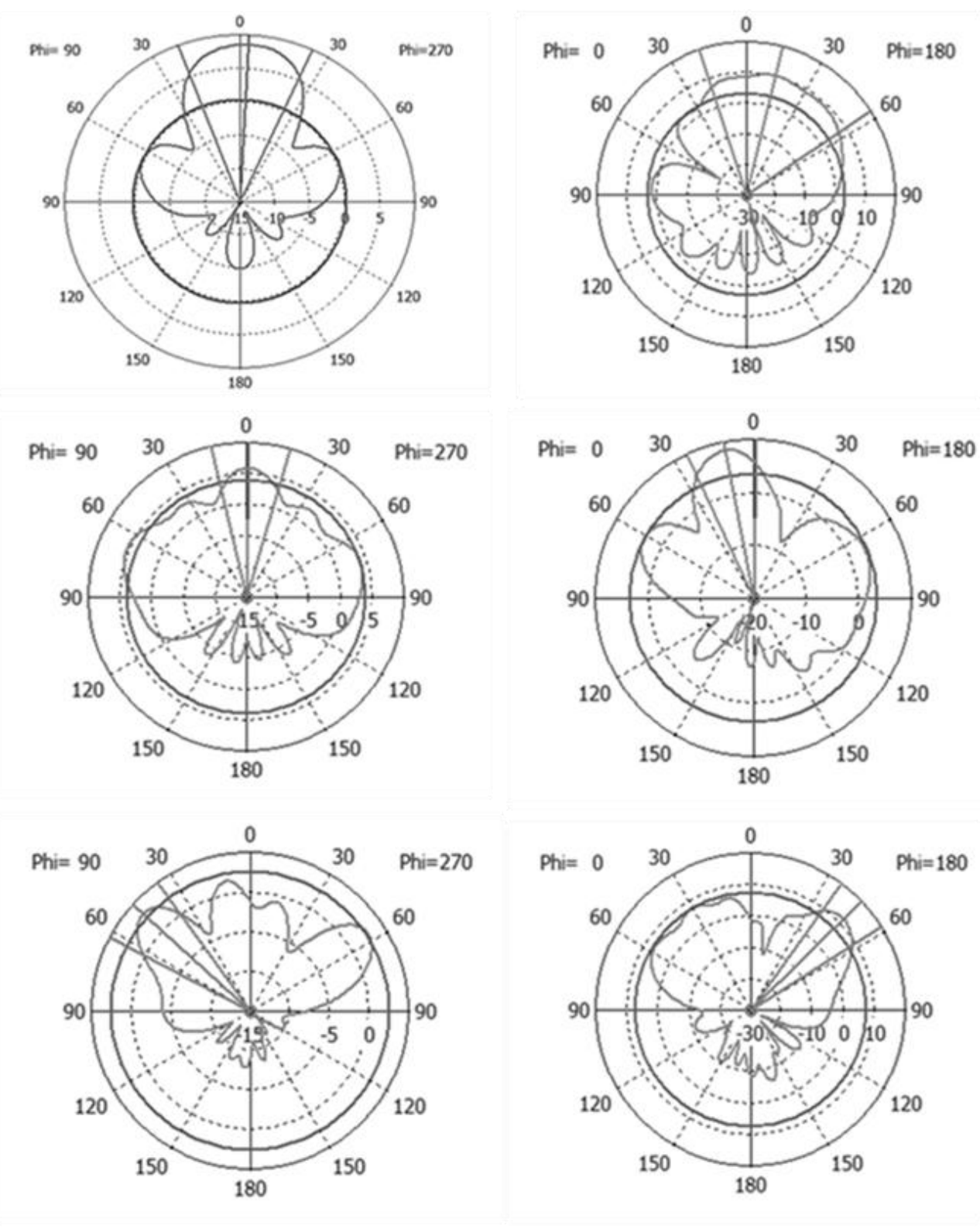

Fig. 5. 3dB radiation patterns of the proposed 1×6 linear antenna array at 28 GHz, showing the radiation cuts in the $\phi = 0°$ and $\phi = 90°$ planes.

The surface current distribution of the proposed 1×6 linear microstrip patch antenna array was analysed at 28 GHz to evaluate the excitation behaviour and electromagnetic energy flow across the structure. As shown in Fig. 6, the current originates from the main feedline and is evenly distributed to all six radiating patches through the series-fed microstrip transmission lines. The highest current density is observed at the edges and feed points of each rectangular patch, which is typical for microstrip antennas operating at resonance. The uniform excitation of all elements confirms that the feeding network is well-designed and that the antenna is properly impedance matched, as also supported by the reflection coefficient and VSWR results. The standing wave pattern observed along the transmission line and patches indicates constructive interference, which contributes to the formation of a directional main lobe in the far-field radiation pattern. The absence of strong current discontinuities or imbalances suggests minimal reflection and loss, ensuring high radiation efficiency. This smooth and symmetrical current distribution validates that the designed antenna array performs as intended, with each element contributing effectively to the overall radiation mechanism.

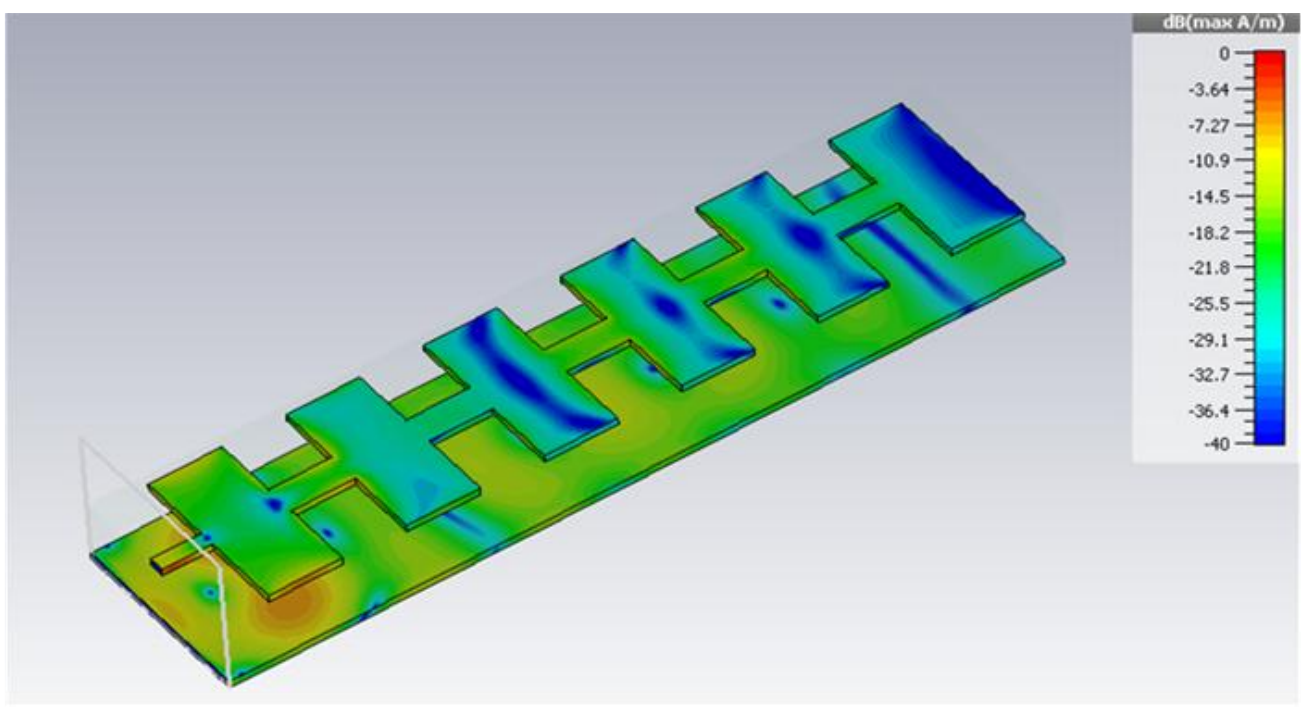

Fig. 6. Surface current distribution of the 1x6 linear antenna array.

## IV. Conclusion

In this work, a compact high-frequency microstrip patch antenna was designed, simulated, fabricated, and analysed for 28 GHz applications, particularly targeting 5G and mm-wave wireless communication systems. The antenna design focused on achieving high gain, proper impedance matching, and directional radiation patterns while maintaining compactness and simplicity in structure. Extensive simulation and measurement results were presented to validate the performance of the antenna in terms of reflection coefficient, VSWR, gain, bandwidth, radiation characteristics, and current distribution. The reflection coefficient analysis revealed a strong resonance at 28.8 GHz, with a minimum value of approximately 16 dB, indicating good impedance matching at the desired frequency. The –10 dB impedance bandwidth extended from 28.3 GHz to 29.6 GHz, providing a usable bandwidth of 2 GHz, which is sufficient to support high-data-rate mm-wave communication systems. The corresponding VSWR remained below 2.0 across the operating band, with a minimum close to 1.1, further confirming efficient power delivery to the antenna with minimal reflections. The simulated peak gain of the antenna is 9.9 dBi, which is suitable for directional point-to-point communication and base station applications in the 5G mm-wave spectrum. The gain remained relatively stable over the entire operational bandwidth, ensuring consistent performance across the frequency band. This characteristic is particularly beneficial in environments where frequency-selective fading can affect link reliability. Radiation pattern analysis demonstrated that the antenna exhibits a directional radiation behaviour, with the main lobe directed at $\theta = 40°$ in the $\phi = 0°$ plane and $\theta = 4°$ in the $\phi = 90°$ plane. The 3 dB angular beamwidths were found to be 86.3° and 82.6°, respectively, which indicates a broad main lobe suitable for covering a significant angular region. The side lobes were well suppressed, with side lobe levels of –3.1 dB and –4.7 dB, minimizing unwanted radiation in other directions and enhancing the antenna's overall directivity. Finally, surface current distribution analysis illustrated a standing wave pattern, with current effectively distributed over the radiating elements. This confirms the antenna's proper mode excitation and supports the observed far-field radiation characteristics. The uniform distribution also ensures that all patches contribute effectively to the radiation mechanism, optimizing performance.

In summary, the proposed antenna meets the essential criteria for high-frequency wireless communication systems, including high gain, broad bandwidth, good impedance matching, and directional radiation. Its compact size and fabrication simplicity make it suitable for practical deployment in 5G mm-wave systems such as base stations, automotive radars, and point-to-point communication links. Future work may involve array integration and beamforming techniques to further enhance coverage and beam agility.